# Space science knowledge in the context of Industry 4.0 and Space 4.0

Roland Walter[1]

## 1. Introduction

Astrophysics deals with the most important questions characterizing the human nature. Where do we come from? What is the future of humanity? How did life begin? Is there life elsewhere? Our understanding of the Universe develops fast, although with detours and surprises. Most of our knowledge, from the source of energy in stars, their assembly in galaxies or the structure and history of the Universe was obtained in one century, largely thanks to the advances of technology. Astronomy has always been and is still a main driver of space science and exploration[2].

Space is among the greatest adventures of mankind, largely the adventure of our generation. Space is dangerous, spectacular, with fascinating phenomena and full of opportunities. The data resulting from explorations, collected in space or on ground, are of great value, as it encodes our knowledge of the Universe, of our past and of our future. Astronomy and space therefore hold a special place in science communication, because of the fundamental questions tackled, providing an exceptional link between science and society. Astronomy and space are also inclusive by nature, all humans sharing the same spaceship.

The United Nation Committee on the Peaceful Uses of Outer Space is now considering space as a driver for sustainable development[3], promoting the use of space data to foster not only collaboration, science and innovation but also inclusion. Providing straightforward interfaces to complex data and their analysis could make the process of generating science available to the society and education at large, a fundamental step to transmit the values of science and to evolve towards a knowledge society, worldwide.

High-energy astrophysics space missions pioneered and demonstrated how powerful legacy data sets can be for generating new discoveries[4], especially when combined with data from different research infrastructures, and analysed in ways that the original researchers could not have anticipated. Nowadays the success of a research infrastructure is not only measured by the answers it provides to important scientific questions but also by the information it makes available to a wide community.

---

[1] Roland Walter (roland.walter@unige.ch) is astrophysicist at the astronomy department of the University of Geneva. He is specialized in the study of high-energy phenomena occurring around compact objects or in diffuse media. He led the development of the data centre for ESA's INTEGRAL mission and of the HEAVENS project and is a member of the Cherenkov Telescope Array consortium.

[2] The 22 states parties to this convention (2010) Article V Activities and Programmes. in: Convention for the establishment of a European Space Agency. ESA Communications, ESA-SP-1317, 7th edition December 2017. Available at http://www.esa.int/About_Us/Law_at_ESA/ESA_Convention. Accessed 19 March 2018.

[3] The Committee on the Peaceful Uses of Outer Space (2018) Draft resolution on space as a driver for sustainable development. United Nations A/AC.105/C.1/L.364, 9 February 2018. Available at https://cms.unov.org/dcpms2/api/finaldocuments?Language=en&Symbol=A/AC.105/C.1/L.364. Accessed 19 March 2018.

[4] White, Nicholas E. (2012) From EXOSAT to the High-Energy Astrophysics Science Archive: X-ray Astronomy Comes of Age. NASA Technical Report GSFC.ABS.7468.2012. Available at https://ntrs.nasa.gov/search.jsp?R=20120016404. Accessed on 19 March 2018.

Space agencies[5], science communities[6], economic and development organisations[7], recognizing that preservation and access to science data are central to their missions, have enforced the policy of having data becoming publicly usable. Since about 40 years the astrophysics community[8] develops standards for sharing data and making them available, mostly within specialized communities.

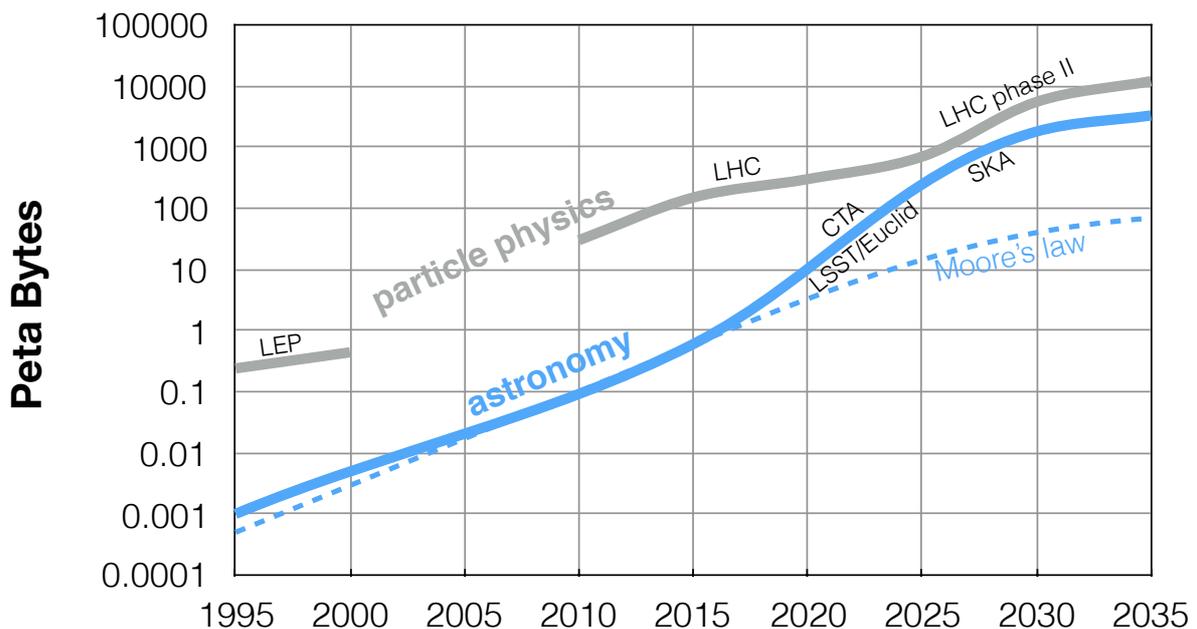

*Figure 1: Evolution of the data volume produced by astronomy experiments compared to that produced by the main particle colliders of the European Organisation for Nuclear Research (CERN). LEP and LHC stands for the Large Electron-Positron collider and the Large Hadron Collider. The acronyms of the astronomical experiments are mentioned in the text.*

The business model of astronomers is however changing as the increase of data complexity and the decrease of the cost of launch services are calling for adaptations.

---

[5] The 22 states parties to this convention (2010) Article III Information and Data. in: Convention for the establishment of a European Space Agency. ESA Communications, ESA-SP-1317, 7th edition December 2017. Available at http://www.esa.int/About_Us/Law_at_ESA/ESA_Convention. Accessed 19 March 2018.

[6] Committee on Issues in the Transborder Flow of Scientific Data (1997) Bits Of Power, Issues in Global Access to Scientific Data. The National Research Council, National Academy Press ISBN 0-309-05635-7. Available at http://www.nap.edu/read/5504/ Accessed 19 March 2018.

[7] The Secretary-General of the OECD (2007) OECD Principles and Guidelines for Access to Research Data from Public Funding. Available at http://www.oecd.org/sti/sci-tech/38500813.pdf). Accessed 19 March 2018.

[8] E.g. the data and documentation commission (https://www.iau.org/science/scientific_bodies/commissions/B2/) and the FITS working group (https://fits.gsfc.nasa.gov/iaufwg/iaufwg.html) of the International Astronomical Union.

The data volume and computing power necessary to operate the upcoming astrophysics research infrastructures will increase by a factor of hundred in the next decade (figure 1). Even when Moore's law is considered, the complexity will increase by a factor of ten. Astrophysicists are entering in the big data era and need an evolution of their business model. A centralisation towards clouds, providing storage, computing, and some standardisation is taking place in the industry and in science[9]. Handling data flows exceeding human insight, and accessing knowledge at a level of abstraction higher than the data themselves requires new tools, in particular artificial intelligence, at the core of the *fourth industrial revolution*[10].

The cost of launch services is reducing drastically, from >10 USD down to about 1 USD per gram brought to low Earth orbit[11], and has the potential to shift the space leadership from the few spacefaring nations to numerous space actors, industries, academia, and citizens. The multiplication of these actors makes their coordination and definition or adoption of standards for data and knowledge management and sharing more challenging. This transition was christened *Space 4.0* at the 16th ministerial council of the European Space Agency[12].

## 2. Evolution of knowledge management

Interfacing the knowledge on the Universe should be considered at different levels: data, analysis, interpretation and knowledge. The volume of details and information decreases along that sequence while the amount of study, reduction and computing increases. The level of abstraction increases from analytic to synthetic and common sense or even wisdom (figure 2).

The first digital archives of astrophysical data were created in the 1970s[3] and developed to provide access to detector and high-level data resulting from a standard data analysis and characterising the observed targets. Data were made accessible first through interfaces dedicated to specific missions. Later, generic portals were adopted along with the world-wide web in the 1990s[13]. Eventually virtual observatory[14,15] protocols were introduced ten years later to help automated data discovery and sharing worldwide. Virtual observatory protocols were also defined to integrate high-level analysis services.

Data mining and knowledge discovery concepts were introduced in the 1990s and made clear that data analysis was an iterative and interactive process as science questions and methods evolve over time. The successive data preparation, modelling and evaluation steps indeed depend on the scientific question tackled and on the user needs.

---

[9] The European Commission (2018) Implementation Roadmap for the European Open Science Cloud. SWD(2018) 83 final. Available at http://ec.europa.eu/research/openscience/pdf/swd_2018_83_f1_staff_working_paper_en.pdf#view=fit&pagemode=none. Accessed 19 March 2018.
[10] The World Economic Forum (2018) Fourth industrial revolution. Available at http://www.weforum.org/agenda/archive/fourth-industrial-revolution. Accessed 19 March 2018.
[11] Falcon heavy can lift 63800 kg to low Earth orbit for 90M$ (http://www.spacex.com/about/capabilities), and 30% less using reusable components (https://twitter.com/elonmusk/status/726559990480150528).
[12] The European Space Agency (2016) What is space 4.0 ? Available at http://www.esa.int/About_Us/Ministerial_Council_2016/What_is_space_4.0. Accessed 19 March 2018.
[13] The World Wide Web Consortium (2000) A Little History of the World Wide Web Web. Available at https://www.w3.org/History.html. Accessed 19 March 2018.
[14] Szalay A.S. and Brunner R.J. (1998) Astronomical Archives of the Future: a Virtual Observatory. Available at https://arxiv.org/pdf/astro-ph/9812335.pdf. Accessed 19 March 2018.
[15] Benvenuti, P. (2001) The Astrophysical Virtual Observatory. STECF Newsletter 29,4, Available at http://adsabs.harvard.edu/abs/2001STECF..29....4B. Accessed 19 March 1018.

Several interfaces appeared from 2005 onwards using such ideas for specific astronomical missions, allowing to analyse data on-the-fly based on user specific parameters. These tools effectively interface users with the analysis rather than with the data and provide improved flexibilities for non-specialists. Some of these systems allow to remotely analyse[16,17] observation data sets while others[18], closer to data mining, start from transformed pre-processed data and allow to quickly generate mission-wide products suitable for various types of analysis and science goals.

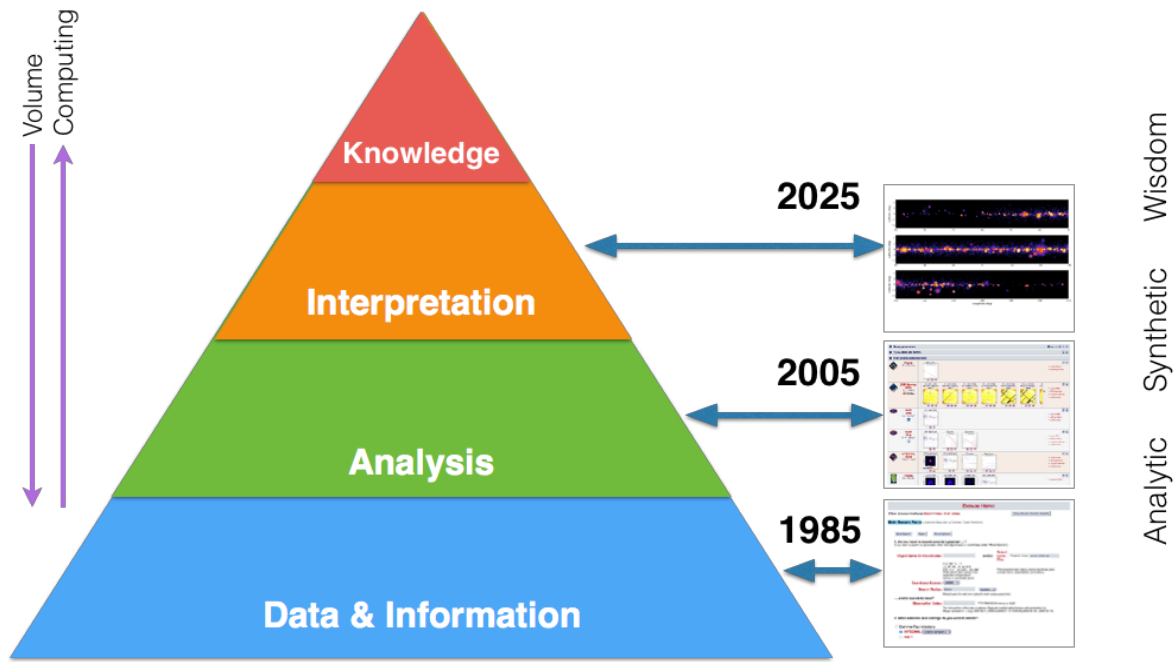

*Figure 2: Evolution of knowledge management in Astronomy.*

As analysis software, specific to an experiment, was rarely maintained over decades, it was practically impossible to repeat analysis, or to extend it to new targets and science goals, especially when an experiment or space mission was over and detailed know-how had vanished. This is a fundamental limit of services, like the virtual observatory, based on pre-calculated archive holdings.

The latter services could provide an analytic access to mission data to everyone if suitable interfaces are provided. Technically, interfaces to such services could be built by extending virtual observatory standards to the needs of remote processing. These services could easily run on virtual machines on clouds, simplifying software maintenance and portability. If designed properly, they can also easily be moved between institutions, if the need arises.

---

[16] Pence, W.D. and Chai P. (2007) The HEASARC HERA system. ADASS XVI. ASPC 376, 554, Available at http://adsabs.harvard.edu/abs/2007ASPC..376..554P. Accessed 19 March 2018.

[17] The UK Swift Science Centre (2014) Swift XRT products generation system. Available at Leicester University http://www.swift.ac.uk/user_objects/. Accessed 19 March 2018.

[18] Walter R., et al. (2010) INTEGRAL in HEAVENS. Available at https://pos.sissa.it/115/162/pdf. Accessed 19 March 2018.

Machine learning algorithms started to have an impact in the early 2000. Deep learning can find interesting results on data sets without the need to write specific software. Supervised machine learning is particularly good to classify data and can be used to search, understand voices, recognize spam e-mails, act as a virtual helpdesk, or drive cars. Deep learning can learn from the multiple translations available in international organizations how to automatically translate texts. It can also be used to recognize galaxies, classify detector events, detect gravitational waves, etc. Countless new applications are developed. Generic deep learning algorithms usually out-perform specific algorithms or even human classification. Deep learning outperforms humans on go, chess, poker and soon for all sort of activities[19]. Machine learning will progress further, generating games, movies on-demand or rewriting this article, in the future.

As training of neural networks is computationally expensive, the industry is developing hardware for accelerated computing, such as Graphical Processing Units (GPUs) or other specialized processors concentrating thousands of computing cores in a single chip. The combination of these software and hardware technologies is transforming our societies in many areas and even "who we are as human beings"[20].

Machine learning algorithms will be used to analyse the data collected by future research experiments studying the Universe, especially these characterised by large and complex data sets. Such techniques can be used for data analysis (e.g. performing data calibration, selection, etc) or data interpretation, providing tools to allow the study of abstractions such as astronomical objects characteristics or their distribution in the Universe. Some examples of the use of neural networks in the analysis or interpretation of astronomical data, which are relevant for the next generation of instruments, are given below.

The Cherenkov Telescope Array[21], is the next generation ground-based observatory for gamma-ray astronomy. It will be built incrementally from 2019 and 2025, will start conducting scientific observations in 2022 and will be made of more than 100 telescopes measuring the signature of very high energy photons and particles producing flashes of blue light when interacting with the Earth atmosphere. These flashes are recorded by cameras imaging the sky up to a billion times per second and generating about 10 Peta Byte of data per year. Separating flashes generated by photons or particles is a key aspect of the data analysis and is performed efficiently by deep neural networks (figure 3).

The space mission Euclid[22] from the European Space Agency and the Large Synoptic Survey Telescope[23] will observe billions of galaxies in the infrared and visible light with the primary goal (among others) to map the geometry of the Universe through the study of weak gravitational lensing, correlations, supernovae etc. Machine learning techniques are very efficient to improve galaxy images, to measure shear of these images created by weak gravitational lensing of dark matter on the line of sight, and to detect strong gravitational lenses[24]. They will be a technique of choice to interpret the Euclid data.

---

[19] Grace, K. et al. (2017) When Will Artificial Intelligence Exceed Human Performance? Available at http://arxiv.org/abs/1705.08807v2. Accessed 19 March 2018.
[20] Schwab, K. (2018) Shaping the Fourth Industrial Revolution. World Economic Forum. Available at https://www.weforum.org/agenda/2018/01/the-urgency-of-shaping-the-fourth-industrial-revolution/. Accessed 19 March 2018.
[21] The Cherenkov Telescope Array Observatory (2016) Exploring the Universe at the Highest Energies. Available at https://www.cta-observatory.org. Accessed on 19 March 2018.
[22] See "Euclid: an ESA mission to map the geometry of the dark Universe" (http://sci.esa.int/euclid), Euclid was approved in 2011 and should be launched in 2021.

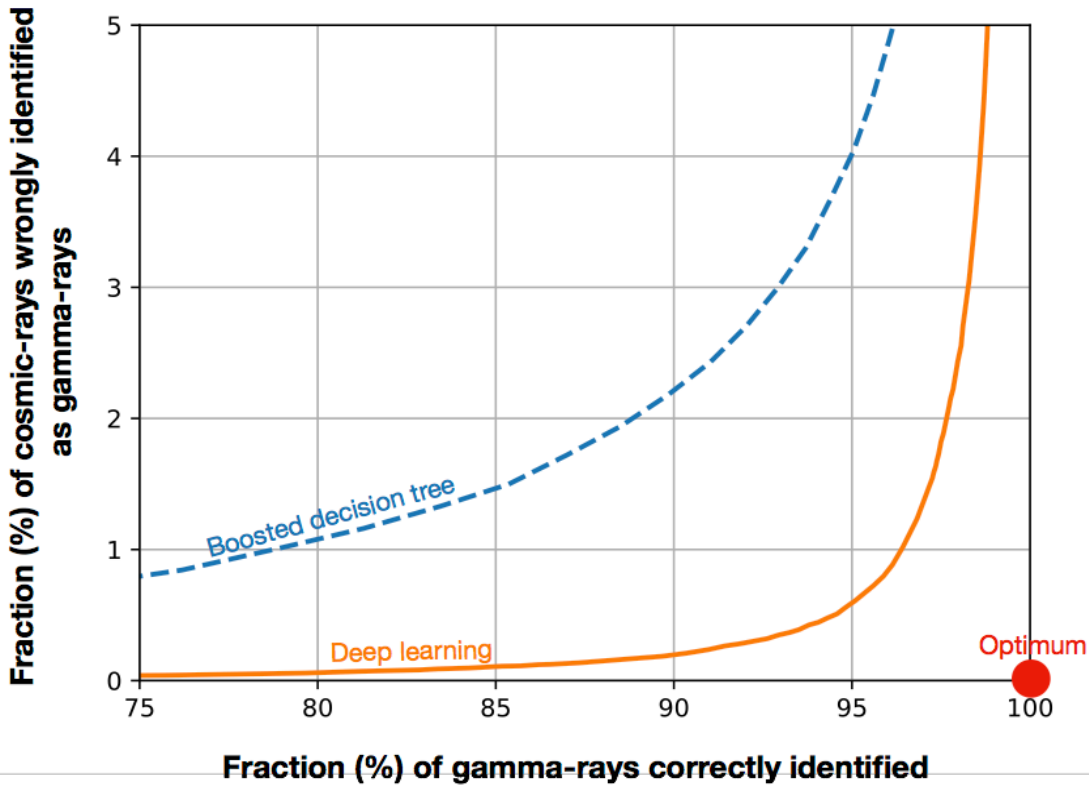

*Figure 3: Comparison of the efficiency of deep learning with classical techniques to separate gamma-ray from cosmic-ray images (for the CTA northern array in the energy band 0.3-1.8 TeV). The red point shows the optimal goal i.e. a perfect separation. The curves show the separation efficiency of particular techniques for continuous values of a selection parameter. The blue (dashed) curve is the state of the art obtained with classical methods (boosted decision trees on parameters extracted from the images). The orange curve was obtained applying deep learning techniques on the same selection of events, providing much improved results and not requiring any prior interpretation of the data, contrasting with classical methods.*

The Square Kilometre Array (SKA)[25] is planned to become the world's largest radio telescope array, with over a square kilometre of collecting area spread across continental scale in Aus-tralia and South Africa. High level, three dimensional images, to be made available daily to the users, will be one Peta Byte in size and each include information for up to a million radio-sources. These volume and source number are too large to be analysed by humans. Neural networks will be mainstream to classify astronomical objects or find specific features[26] e.g. to give access to statistical information on the ensemble of detected sources.

---

[23] See "Large Synoptic Survey Telescope" (https://www.lsst.org). LSST is being built and its first light is planned in 2020.
[24] E.g. "Generative Adversarial Networks recover features in astrophysical images of galaxies beyond the deconvolution limit" (https://arxiv.org/abs/1702.00403) & "Deep Convolutional Neural Networks as strong gravitational lens detectors" (https://arxiv.org/abs/1705.07132) & "Hopfield Neural Network deconvolution for weak lensing measurement" (https://arxiv.org/abs/1411.3193).
[25] The SKA Organisation (2018) The Square Kilometre Array, exploring the Universe with the world's largest radio telescope. Available at https://www.skatelescope.org. Accessed 19 March 2018.

These experiments, observing a large fraction of the sky, will produce large data flows and provide information for many additional scientific studies than their primary goals. Deep learning will be very efficient to handle data sets exceeding human grasp. As the value of data lies in their use, offering interfaces to interpret these data is fundamental. The idea is to shift from the current paradigm (obtaining data) to allow users to directly build more abstract and deeper understanding. This can be done in various contexts for instance by selecting (allowing users to specify learning samples), cleaning (e.g. using generative adversarial networks), or defining measurable and extracting statistics on them from many data sets.

Such interfaces are not yet available and will be very helpful to scientists and could be extended to other users. Neural networks are also at the base of natural language interfaces which could be adapted to the handling of statistical and scientific data[27]. Even the public could get access to that knowledge if suitable interfaces to interpretation (catalogues, source characterisations, models) are made available. Data2Dome[28] (supported by the European Southern Observatory), which displays astronomical data and models in planetariums in real time, or the ESASky web interface[29] (supported by the European Space Agency), providing access to data from space astronomy missions on the entire sky, are very interesting initiatives in this context and could be extended to the needs of scientific interpretation.

## 3. Challenges

Even in astronomy, a subject appealing to the public, making data and knowledge available and used widely is a challenge and will be even more so in the future because of the complexity and size of the data and of the multiplication of providers. This is also true in technologically advanced economies, which may face a rift between people benefitting or challenged by the implications of artificial intelligence. People need the opportunity to be exposed to artificial intelligence techniques, to understand one of the driving force of the economy and the responsibilities and ethical questions[30] associated.

The current paradigm (providing access to data) suffers from several difficulties which make data hard to use beyond the circle of specialists. The expansion of that circle will benefit from resurfacing the data (a significant fraction of the data is not publicly accessible) and analysis tools (which are often not usable, not accessible, not documented, too complex, or, even worse, lost or not maintained). Ways to improve the current paradigm are discussed elsewhere in this volume[31]. If these actions will be beneficial for the scientific community, they may not be good enough for the society, effectively maintaining a digital divide between specialists and outsiders.

---

[26] Aniyan, A. and Thorat, K. (2017) Classifying radio galaxies with convolution neural network. ApjS 230, 20. Available at https://arxiv.org/abs/1705.03413. Accessed 19 March 2018.

[27] Neelakantan A. et al. (2017) Learning a Natural Language Interface with Neural Programmer. ICLR 2017. Available at https://arxiv.org/abs/1611.08945. Accessed 19 March 2018.

[28] IPS and Data2Dome (2017) Data2Dome bringing together astronomy data providers, science centre professionals and software vendors to advance the state of the art in big data visualisation. Available at http://www.data2dome.org. Accessed 19 March 2018.

[29] The European Space Agency (2018) ESAsky. Available at http://sky.esa.int/. Accessed 19 March 2018.

[30] Brundage M., et al. (2018) The Malicious Use of Artificial Intelligence: Forecasting, Prevention, and Mitigation. Available at https://arxiv.org/abs/1802.07228. Accessed 19 March 2018.

If one wishes the data to be used beyond the circle of specialists, to transmit the values of science, for inclusion and sustainable development, improvements are needed at different levels, from the use of the underlying infrastructure to more abstract user interfaces. Accompanying the shift or paradigm requires actions: a few are suggested below.

*The knowledge on the Universe is a universal heritage*: The data and the knowledge collected on the Universe should be considered as a heritage of the humanity, available and preserved for all the intelligent species of the Universe.

*Science clouds*: Data exploitation will require increasing computing power and possibly less specific software if deep learning continue to develop. Scientists need access to centralized and cost effective computing infrastructures, especially GPUs, and to standards for the provision of services. The European Commission is proposing to interconnect public and private data infrastructures in Europe to develop a European Open Science Cloud[32]. For this purpose, a set of standards and infrastructures to support science are being created. If successful, these standards should also be the base of the data and software resurfacing efforts mentioned above.

*Affordable computing*: Opening data exploitation to the science community at large (beyond the circle of the specialists), to education and the public requires, at some level, freely available computing. The public should be allowed to interface with the data and their interpretation, using free services as a return for public funding. Scientists not belonging to an experiment should manage to access sufficient resources to allow using the data at all levels, if required.

*Interfacing analysis, interpretation and knowledge*: Interfacing data does not provide the level of abstraction required by non-specialists. Services can be built to provide higher level interfaces and to make the data and the products of their analysis available to modern analysis techniques. Interfaces could be built up to a level where they can provide information, user driven analysis and knowledge usable by scientists, education and the public.

*Partnership between actors*: In a world where space data providers will be more independent and numerous, the question of standardisation of data and software becomes more challenging. Open science clouds can help this process at least for publicly funded experiments. The integration of privately funded research experiments might be helped by public-private partnership for data exploitation, and at least by the availability of open standards and codes of conduct.

---

[31] Giommi, P. (2018) The Open Universe Initiative. This volume.
[32] The undersigning stakeholders (2017) European Open Science Cloud Declaration. Available at https://ec.europa.eu/research/openscience/pdf/eosc_declaration.pdf#view=fit&pagemode=none. Accessed 19 March 2019.

*Mitigating the artificial intelligence divide*: artificial intelligence will have a large impact on science and societies, acting positively and negatively. People in general and education need to be confronted to these techniques. Providing proper interfaces could allow at least students at various levels to experiment with modern analysis techniques on astrophysics knowledge, especially that derived from space experiments.

The United Nations Office for Outer Space Affairs (UNOOSA), promoting the cooperation between states and space actors, can play an important role towards a consensus for instance in establishing a code of conduct on data handling, in partnership with the scientific community. It could also raise the importance of establishing and coordinating infrastructures capable of inclusion at the level of the states, international organisations and the economy. Finally, it could take the opportunity of UNISPACE+50 to promote the data and knowledge on the Universe as an intangible world cultural heritage[33].

It was a pleasure to participate to the conference on Space 2030 and Space 4.0 synergies for capacity building in the XXIst Century, co-organized by the European Space Policy Institute and by the UNOOSA on the 3rd of February 2018, to hear points of view on topics from a broad perspective and to share the above considerations.

*Acknowledgements: Figure 3 was created with the help of a grant from the Centre for Advanced Modelling Science at the University of Geneva, computing at the Swiss National Supercomputing Centre and advice from the Swiss Data Science Centre. Figure 1 includes materials from http://www.isdc.unige.ch and https://www.cta-observatory.org.*

---

[33] The 176 states parties to the convention (2003) Convention for the Safeguarding of the Intangible Cultural Heritage. Available at https://ich.unesco.org/en/convention. Accessed 19 March 2018.